\documentclass[runningheads]{llncs}

\usepackage[T1]{fontenc}
\usepackage{amsmath}
\usepackage{amssymb}
\usepackage{graphicx}
\usepackage{xcolor}
\usepackage{colortbl}
\usepackage{array}
\usepackage{multirow}
\usepackage{adjustbox}
\usepackage{booktabs}

\begin{document}

\definecolor{grey-light}{RGB}{235, 235, 235}
\definecolor{grey-medium}{RGB}{185, 185, 185}
\definecolor{grey-dark}{RGB}{135, 135, 135}

\title{Post-Processing in Local Differential Privacy: An Extensive Evaluation and Benchmark Platform}

\titlerunning{Post-Processing in Local Differential Privacy}

\author{Alireza Khodaie \and
Berkay Kemal Balioglu \and
M. Emre Gursoy}
\authorrunning{A. Khodaie et al.}
%
\institute{Department of Computer Engineering, Koç University, Istanbul, Turkiye
\email{\{akhodaie22, bbalioglu23, emregursoy\}@ku.edu.tr}}
\maketitle 

\begin{abstract}

Local differential privacy (LDP) has recently gained prominence as a powerful paradigm for collecting and analyzing sensitive data from users' devices. However, the inherent perturbation added by LDP protocols reduces the utility of the collected data. To mitigate this issue, several post-processing (PP) methods have been developed. Yet, the comparative performance of PP methods under diverse settings remains underexplored. In this paper, we present an extensive benchmark comprising 6 popular LDP protocols, 7 PP methods, 4 utility metrics, and 6 datasets to evaluate the behaviors and optimality of PP methods under diverse conditions. Through extensive experiments, we show that while PP can substantially improve utility when the privacy budget is small (i.e., strict privacy), its benefit diminishes as the privacy budget grows. Moreover, our findings reveal that the optimal PP method depends on multiple factors, including the choice of LDP protocol, privacy budget, data characteristics (such as distribution and domain size), and the specific utility metric. To advance research in this area and assist practitioners in identifying the most suitable PP method for their setting, we introduce LDP$^3$, an open-source benchmark platform. LDP$^3$ contains all methods used in our experimental analysis, and it is designed in a modular, extensible, and multi-threaded way for future use and development.

\end{abstract}

\section{Introduction} \label{sec:introduction}

Over the last few years, local differential privacy (LDP) has become a popular notion for privacy-preserving data collection and analysis. In LDP, users perturb their sensitive data locally and then send it to a data collector. The added perturbation disables the data collector from making inferences regarding any particular user, but the data collector can still estimate population-level statistics from aggregates of perturbed data. This feature of LDP makes it well-suited for data collection in decentralized settings. Consequently, LDP has seen widespread adoption in various research fields \cite{cormode2018privacy,du2023ldptrace,gursoy2019secure,kim2019collecting,yang2023local} as well as industry products, e.g., Google Chrome, Apple iOS, and Microsoft Windows \cite{ding2017collecting,erlingsson2014rappor,thakurta2017emoji}.

With the emerging popularity of LDP, numerous LDP protocols have been developed in the literature \cite{cormode2021frequency,gursoy2022adversarial,wang2017locally}. However, since these protocols inherently introduce perturbation to ensure privacy, the population-level statistics estimated by the data collector become noisy, leading to utility loss. To reduce utility loss, several post-processing (PP) methods were proposed \cite{jia2019calibrate,wang2020locally}, e.g., Base-Pos, Norm, Norm-Cut, Norm-Sub, Norm-Mul, Power, and Power-NS. In essence, PP methods take as input the estimation results and process them in different ways to improve utility. On the other hand, although PP methods have been utilized in specific contexts or tested with specific protocols, to the best of our knowledge, there is no benchmark or detailed analysis of how different PP methods behave comparatively under diverse settings, e.g., under different combinations of LDP protocols, data characteristics, and utility metrics. 

In this paper, we create an extensive setup containing 6 LDP protocols, 7 PP methods, 4 utility metrics, and 6 datasets to benchmark the behaviors and optimality of PP methods under diverse settings. We perform an experimental analysis using our benchmark setup and report (a subset) of our results in the paper. Our results and analyses yield several interesting findings. First, we find that PP methods indeed yield substantial utility benefits when the privacy budget $\varepsilon$ is small (strict privacy), but their utility results converge as $\varepsilon$ becomes larger, and their benefits decrease. Second, we find that the best-performing PP method (in terms of utility improvement) is not static, but rather, it depends on a variety of factors including the LDP protocol, $\varepsilon$ budget, data characteristics (e.g., statistical distribution, domain size), and utility metric. Third, we find that the best-performing PP method is more consistent when $\varepsilon$ is small, but it becomes more inconsistent as $\varepsilon$ increases.

Next, considering there are multiple factors that influence the performance of PP methods and the choice of optimal PP method, we ask the following question: Can we design a practical system that helps a practitioner choose the best PP method for their setting, i.e., data, protocol, budget, and utility metric? To address this question, we design and develop LDP$^3$ (pronounced LDP-Cube): an open-source benchmark platform containing the implementations of all protocols, PP methods, utility metrics, and datasets used in our benchmark. LDP$^3$ is designed in a modular and extensible way so that new protocols, PP methods, utility metrics, etc.~can be added in the future. Furthermore, it supports multi-threaded execution to reduce execution times so that researchers and practitioners can perform fast benchmarking of PP methods.

\vspace{5pt}
\noindent
\textbf{Contributions.} In short, in this paper we make three main contributions:
\begin{itemize}
    \item We create a benchmark setup with 6 LDP protocols, 7 PP methods, 4 utility metrics, and 6 datasets to analyze the behaviors and optimality of PP methods under diverse settings.
    \item We perform an experimental analysis and show that the best-performing PP method depends on a variety of factors, including the LDP protocol, $\varepsilon$ budget, data characteristics (e.g., statistical distribution, domain size), and utility metric. We show that when these factors are different, the best-performing PP method can also be different.
    \item To advance research in this area and to assist researchers and practitioners in selecting the best PP method for their setting, we develop an open-source benchmark platform called LDP$^3$ which contains all protocols, PP methods, metrics, and datasets used in our analysis.
\end{itemize}

\noindent
\textbf{Related work and main differences.} While there exist a few past works in benchmarking LDP protocols or developing open-source LDP platforms, our work has some key differences. Earlier works such as \cite{wang2017locally} and \cite{wang2020locally} focus solely on protocols without post-processing or evaluate post-processing only on a single protocol. \textsc{Pure-LDP} proposed in \cite{cormode2021frequency} contains a subset of the LDP protocols and PP methods that exist in LDP$^3$. The \textsc{Multi-Freq-LDP} package was proposed in \cite{Arcolezi2022}; however, its focus is on multi-dimensional and longitudinal frequency estimation \cite{arcolezi2021random,arcolezi2022improving} rather than post-processing. \textsc{LDPLens} was proposed in \cite{gursoy2022adversarial}; however, it is focused on protocols' adversarial analysis and does not contain post-processing. Overall, our work differs in terms of its emphasis on post-processing as well as the novel experimental analysis and take-away messages concerning post-processing methods. 

\vspace{5pt}
\noindent
\textbf{Paper organization.} The rest of the paper is organized as follows. In Section \ref{sec:LDP-PP}, we provide background information related to LDP, LDP protocols, and PP methods. In Section \ref{sec:ExperimentSetup}, we provide information regarding our experiment setup. In Section \ref{sec:Experiments}, we present and discuss our experiment results; additionally, we summarize the main take-away messages from our experimental analysis. In Section \ref{sec:LDPCube}, we describe the design and implementation of LDP$^3$. Finally, Section \ref{sec:Conclusion} concludes the paper.

\section{LDP and Post-Processing Methods} \label{sec:LDP-PP}

\subsection{Local Differential Privacy (LDP)}

Local differential privacy (LDP) is a popular notion for collecting data while protecting individuals' privacy. In a typical LDP setup, there are multiple users (clients) and a server (data collector). We denote the user population by $\mathcal{P}$, a user by $u \in \mathcal{P}$, and the user's true value by $v_u \in \mathcal{D}$, where $\mathcal{D}$ is the domain of allowed values. To satisfy LDP, each user encodes and perturbs their $v_u$ locally on their device using a randomized algorithm $\psi$. The user then sends the perturbed value to the server. After gathering data from all users, the server applies estimation methods to compute statistics about the overall population $\mathcal{P}$. However, since each user's data is perturbed using $\psi$, the server cannot infer exact information about any particular user's $v_u$. 

\begin{definition}[$\varepsilon$-LDP] \label{def:LDP}
A randomized mechanism $\psi$ satisfies $\varepsilon$-LDP if and only if, for any two values $v_1, v_2 \in \mathcal{D}$:
\begin{equation} \label{eq:LDPeqn}
\forall y \in Range(\psi): ~~~~ \frac{\text{Pr}[\psi(v_1) = y]}{\text{Pr}[\psi(v_2) = y]} \leq e^{\varepsilon}
\end{equation}
where $Range(\psi)$ denotes the set of all possible outputs of $\psi$. 
\end{definition}

$\varepsilon$-LDP guarantees that, given the perturbed value $y$, the server cannot distinguish whether the user’s true value was $v_1$ or $v_2$ with a probability ratio greater than $e^\varepsilon$. The strength of privacy protection is determined by the parameter $\varepsilon$, often referred to as the \textit{privacy budget}, where smaller $\varepsilon$ indicates stronger privacy.

The popularity of LDP has led to the development of various LDP protocols \cite{cormode2021frequency,gursoy2022adversarial,wang2017locally}. These protocols are often used as building blocks in more complex data analysis tasks and downstream applications. An LDP protocol can be characterized by two main components: (i) user-side encoding and perturbation to satisfy LDP, and (ii) server-side aggregation and estimation to recover population-level statistics. In this paper, we utilize six state-of-the-art LDP protocols: GRR, BLH, OLH, RAPPOR, OUE, and SS. Due to the page limit, we do not provide technical descriptions of each protocol; instead, we refer the readers to \cite{gursoy2022adversarial,wang2017locally} which explain the user-side and server-side processes of the protocols in detail.

\subsection{Post-Processing (PP) Methods} \label{sec:PPmethods}

For $v \in \mathcal{D}$, let $f(v)$ denote the true occurrence frequency of $v$ in the user population $\mathcal{P}$. After server-side estimation, the server recovers estimated frequencies $\hat{f}(v)$. However, $\hat{f}(v)$ is typically different from $f(v)$ because of LDP's perturbation, which causes utility loss (error) in downstream analyses and tasks. To improve utility, server-side post-processing (PP) methods have been proposed \cite{jia2019calibrate,wang2020locally} which take as input $\hat{f}(v)$ and produce post-processed frequencies $\tilde{f}(v)$. Different PP methods result in varying trade-offs between utility increase and bias. In this paper, we use the following 7 state-of-the-art PP methods. 

\textbf{Base-Pos:} Frequencies must be non-negative by definition; however, due to the perturbation in LDP, $\hat{f}(v)$ may be negative for some $v \in \mathcal{D}$. Base-Pos addresses this problem by converting all negative estimations to 0. 
\begin{equation}
    \forall v \in \mathcal{D}:
\tilde{f}(v) =
\begin{cases} 
\hat{f}(v) & \text{if } \hat{f}(v) \geq 0 \\
0 & \text{otherwise}
\end{cases}
\end{equation}

\textbf{Norm:} The sum of frequencies across all $v \in \mathcal{D}$ should equal 1; however, this may not hold due to the perturbation in LDP. Norm addresses this problem by adding a constant $\sigma$ to each frequency so that the sum will equal 1.
\begin{equation}
    \forall v \in \mathcal{D}:
\tilde{f}(v) = \hat{f}(v) + \sigma, \quad \text{such that} \quad \sum_{v \in \mathcal{D}} \tilde{f}(v) = 1
\end{equation}

\textbf{Norm-Cut:} Norm-Cut converts negative and small positive frequencies to 0, and it also ensures that the sum of frequencies equals 1. That is:
\begin{equation}
    \forall v \in \mathcal{D}:
\tilde{f}(v) =
\begin{cases}
0 & \text{if } \hat{f}(v) \leq \theta \\
\hat{f}(v) & \text{if } \hat{f}(v) > \theta
\end{cases}
\end{equation}
where $\theta$ is a threshold value. The value of $\theta$ is chosen such that $\sum_{v \in \mathcal{D}} \tilde{f}(v) = 1$ is ensured. 

\textbf{Norm-Sub:} Norm-Sub converts negative frequencies to 0. Then, it adds a constant $\delta$ to the frequencies to ensure that the sum of frequencies equals 1.
\begin{equation}
    \forall v \in \mathcal{D}:
\tilde{f}(v) =
\begin{cases}
0 & \text{if } \hat{f}(v) < 0 \\
\hat{f}(v) + \delta & \text{if } \hat{f}(v) \geq 0
\end{cases}
\end{equation}
Here, the value of $\delta$ is chosen such that $\sum_{v \in \mathcal{D}} \tilde{f}(v) = 1$ is ensured. 

\textbf{Norm-Mul:} Norm-Mul converts negative frequencies to 0. Then, instead of an additive factor, it uses a multiplicative factor to the remaining frequencies so that their sum becomes 1.
\begin{equation}
    \forall v \in \mathcal{D}:
\tilde{f}(v) =
\begin{cases}
0 & \text{if } \hat{f}(v) < 0 \\
\alpha \hat{f}(v) & \text{if } \hat{f}(v) \geq 0
\end{cases}
\end{equation}
Here, $\alpha$ is the multiplicative factor. Its value is chosen such that $\sum_{v \in \mathcal{D}} \tilde{f}(v) = 1$ is ensured.

\textbf{Power:} The rationale of Power is that many real-world datasets follow a statistical distribution such as a Gaussian or power law distribution. Therefore, Power fits a distribution to the estimated frequencies and aims to minimize the expected square error. For example, with $\tilde{f}(v) \sim P$ where $P$ is a power law distribution, the goal is to minimize:
\begin{equation}
    \min_{P} ~ \mathbb{E}\left[ \sum_{v \in \mathcal{D}} \left( \hat{f}(v) - \tilde{f}(v) \right)^2 \right]
\end{equation}

\textbf{Power-NS:} Power-NS first applies Power and then uses Norm-Sub on Power's outputs to obtain the final result.

\section{Benchmark Setup} \label{sec:ExperimentSetup}

We used Python to implement all LDP protocols, PP methods, and experiments. Each experiment was conducted 20 times for statistical significance. 

\subsection{Datasets} \label{sec:Datasets}

We used three real-world datasets (BMS-POS, Kosarak, Adult) and three synthetically generated datasets (Zipfian, Gaussian, Uniform) that are also commonly used in the DP and LDP literatures for experimentation.

\textbf{Adult} is a well-known dataset in the privacy literature, consisting of 45,222 records and 14 features. We downloaded it from the UCI ML Repository\footnote{\url{https://archive.ics.uci.edu/dataset/2/adult}}. The features relate to individuals' census information such as age, education, workclass, marital status, race, and gender. In our experiments, we used \textit{age} values as users' true values. Ages in the dataset ranged between 17 and 90; therefore, we have $|\mathcal{D}| = 74$.

\textbf{Kosarak} contains click-stream data of a Hungarian online news portal. We downloaded it from the SPMF Dataset Repository\footnote{\url{https://www.philippe-fournier-viger.com/spmf/index.php?link=datasets.php}}. Each row in the dataset corresponds to one user's URL visit sequence. Due to many URLs having few occurrences (e.g., one or two), we pre-processed the dataset by identifying the top 128 most visited URLs and removed the rest, i.e., $|\mathcal{D}| = 128$. For users who had more than one URL in their resulting stream, the most frequently occurring URL in their stream was picked as their $v_u$. 

\textbf{BMS-POS} includes market basket sales data from a major electronics retailer, consisting of 515,596 transactions and 1,657 unique items sold. We downloaded it from the public Github repository\footnote{\url{https://github.com/cpearce/HARM/blob/master/datasets/BMS-POS.csv}} and pre-processed it in a similar fashion to Kosarak by keeping only the top 256 most frequently purchased items. 

\textbf{Gaussian:} We synthetically generated multiple Gaussian datasets in which users' true values are sampled from a Gaussian distribution with mean $\mu$ = 50 and standard deviation $\sigma$ = 1, 5, 10, and 20. In all Gaussian datasets, the domain size is fixed to $|\mathcal{D}|$ = 100 and the population size is fixed to $|\mathcal{P}|$ = 100,000. 

\textbf{Zipfian:} We generated Zipfian datasets in which users' true values are sampled from a Zipfian distribution. The Zipfian distribution is characterized by a skewness parameter $s$. In our datasets, we fixed $s$ = 1.5 and $|\mathcal{P}|$ = 1,000,000, and varied the domain sizes $|\mathcal{D}|$ = 32, 64, 128, 256, 512, 1024, and 2048. 

\textbf{Uniform:} Finally, we generated Uniform datasets in which users' true value distribution follows a Uniform distribution. We used $\mathcal{|P|} = 100,000$ by default and the domain size $|\mathcal{D}|$ is varied.

\subsection{Utility Metrics} \label{sec:UtilityMetrics}

We utilized four utility metrics to measure the errors between the true frequencies $f(v)$ and the post-processed frequencies $\tilde{f}(v)$: $\ell_1$ distance, $\ell_2$ distance, Kullback-Leibler Divergence (KL), and Earth Mover's Distance (EMD). Unless otherwise stated, $\ell_1$ distance is the default utility metric used when reporting results.

\textbf{$\ell_1$ distance}, also known as Manhattan distance, measures error using the absolute value norm. It is defined as:
\begin{equation}
    \text{$\ell_1$ distance} = \sum_{v \in \mathcal{D}} \left| \tilde{f}(v) - f(v)  \right|
\end{equation}

\textbf{$\ell_2$ distance}, also known as Euclidean distance, measures error using the squared norm. It is defined as:
\begin{equation}
    \text{$\ell_2$ distance} = \sqrt{ \sum_{v \in \mathcal{D}} \left( \tilde{f}(v) - f(v) \right)^2 }
\end{equation}

\textbf{Kullback-Leibler divergence} is used to measure how much one probability distribution diverges from another. In this context, $f(v)$ and $\tilde{f}(v)$ are treated as probability mass functions, and KL-divergence is defined as:
\begin{equation}
    \text{KL-divergence} = \sum_{v \in \mathcal{D}} f(v) \log \left( \frac{f(v)}{\tilde{f}(v)} \right)
\end{equation}

\textbf{Earth Mover's Distance (EMD)} measures the minimum cost of transforming one distribution into another. It can be interpreted as the amount of work required to transform the original frequency distribution $f(v)$ into the post-processed distribution $\tilde{f}(v)$. Mathematically, it is defined as:
\begin{equation}
\text{EMD} = \min_{\pi} \sum_{v_1, v_2 \in \mathcal{D}} \pi(v_1, v_2) \cdot |v_1 - v_2|
\end{equation}
where $\pi(v_1, v_2)$ represents the amount of mass moved from $v_1$ to $v_2$, and $|v_1 - v_2|$ is the distance between $v_1$ and $v_2$. 

\section{Results and Discussion} \label{sec:Experiments}

\subsection{Impacts of LDP Protocols and Privacy Budgets}

\begin{figure}[!t]
    \centering
    \begin{minipage}[b]{0.32\textwidth}
        \centering
        \includegraphics[width=\textwidth]{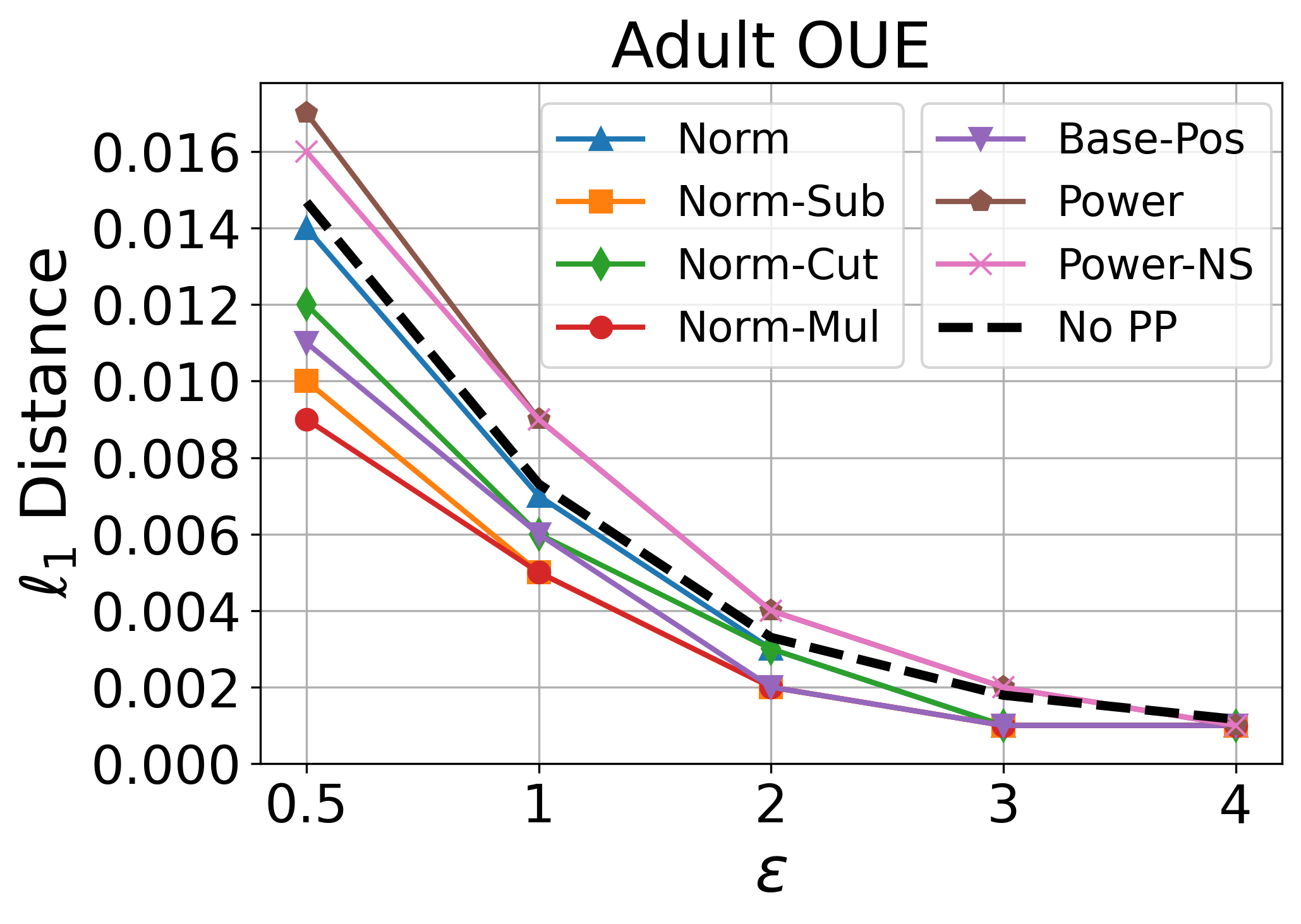}
    \end{minipage}
    \begin{minipage}[b]{0.32\textwidth}
        \centering
        \includegraphics[width=\textwidth]{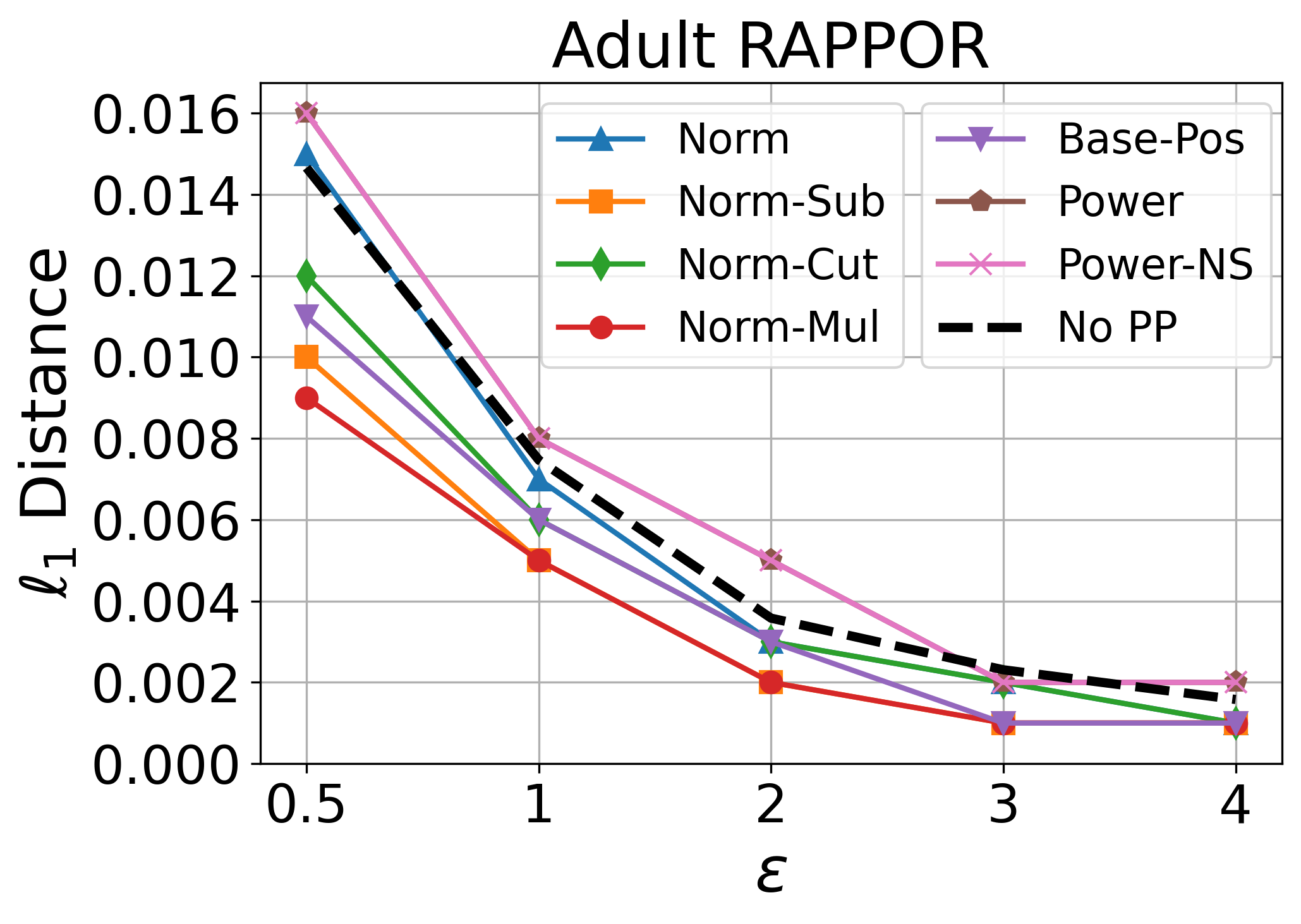}
    \end{minipage}
    \begin{minipage}[b]{0.32\textwidth}
        \centering
        \includegraphics[width=\textwidth]{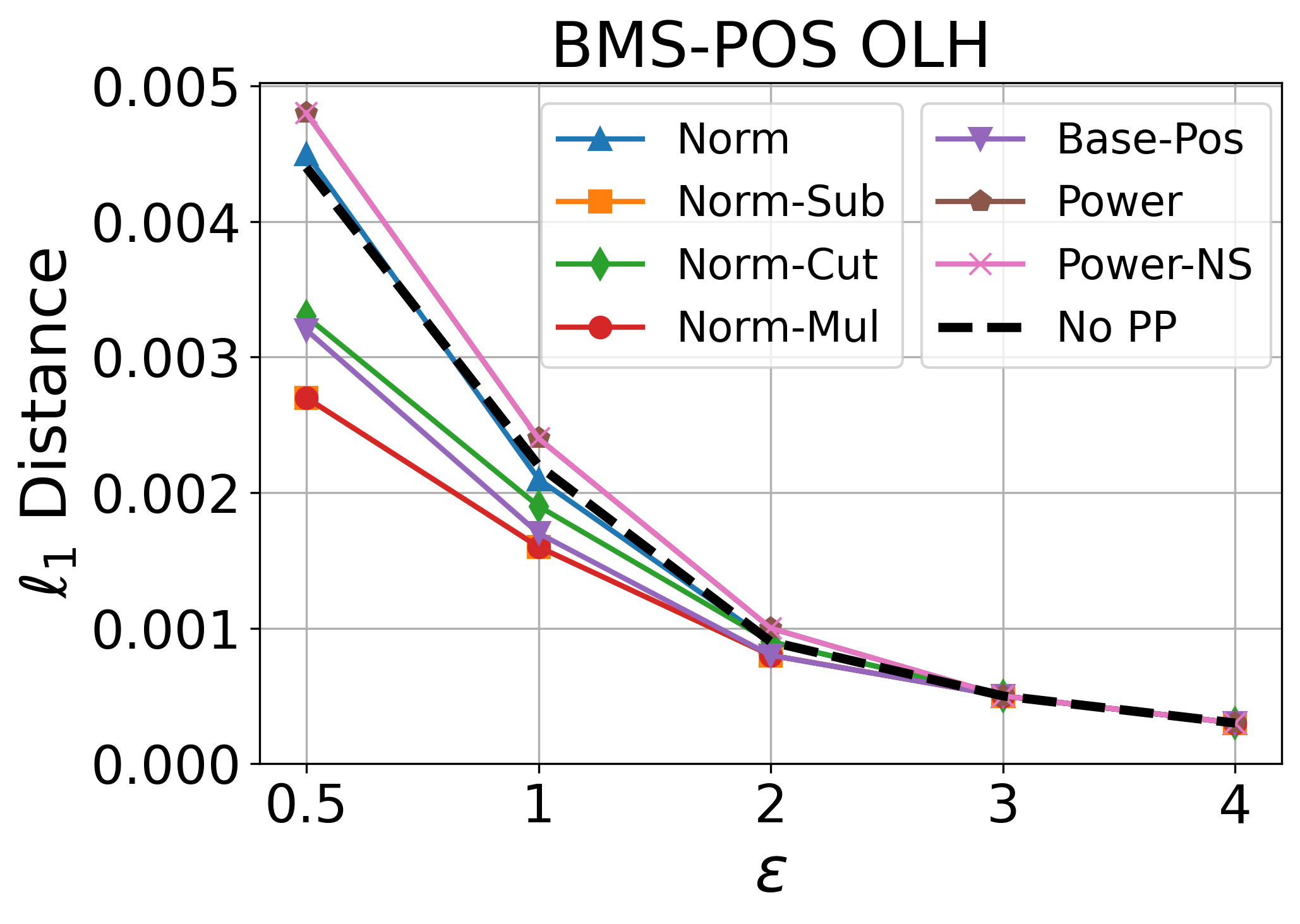}
    \end{minipage}
    \begin{minipage}[b]{0.32\textwidth}
        \centering
        \includegraphics[width=\textwidth]{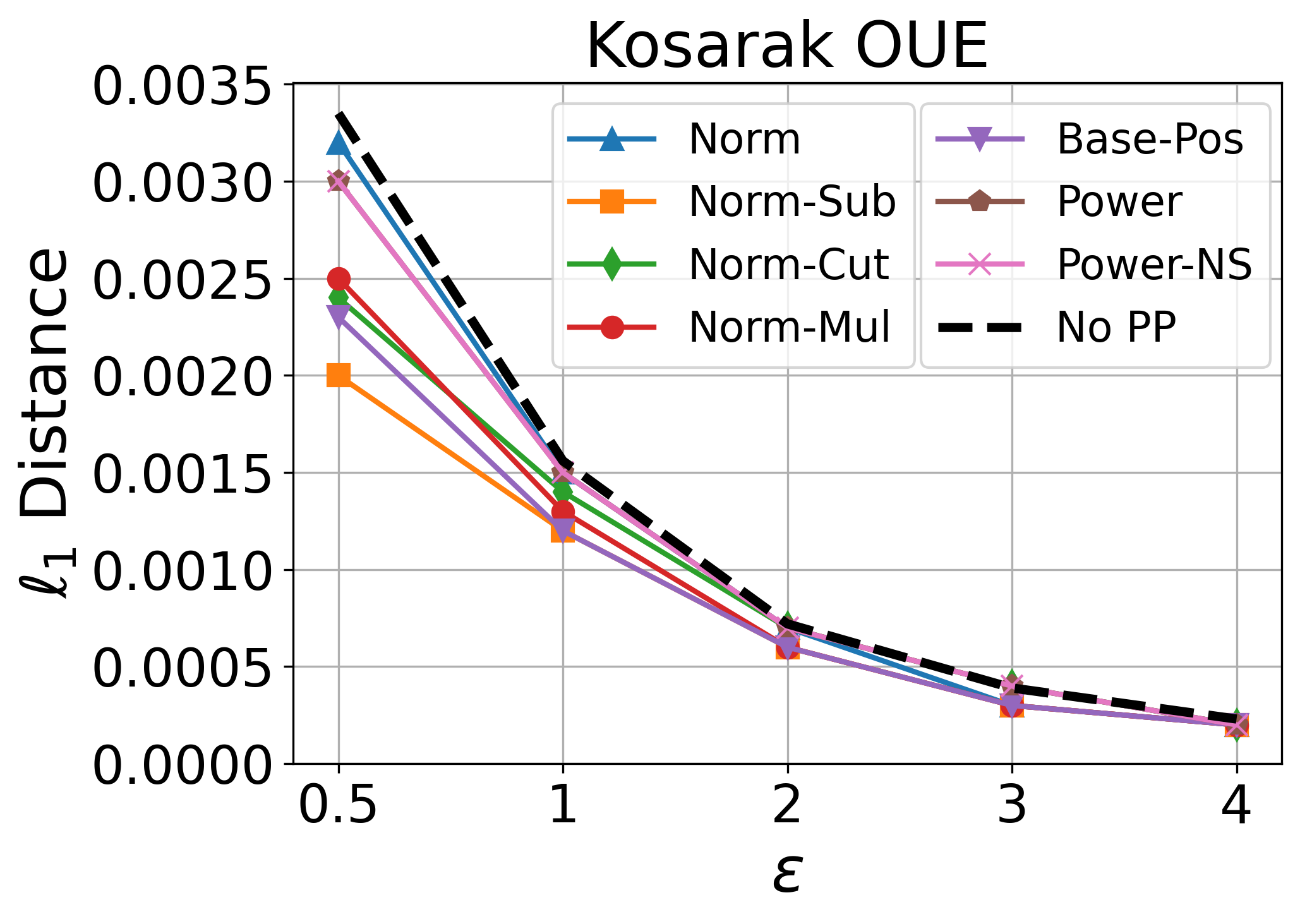}
    \end{minipage}
    \begin{minipage}[b]{0.32\textwidth}
        \centering
        \includegraphics[width=\textwidth]{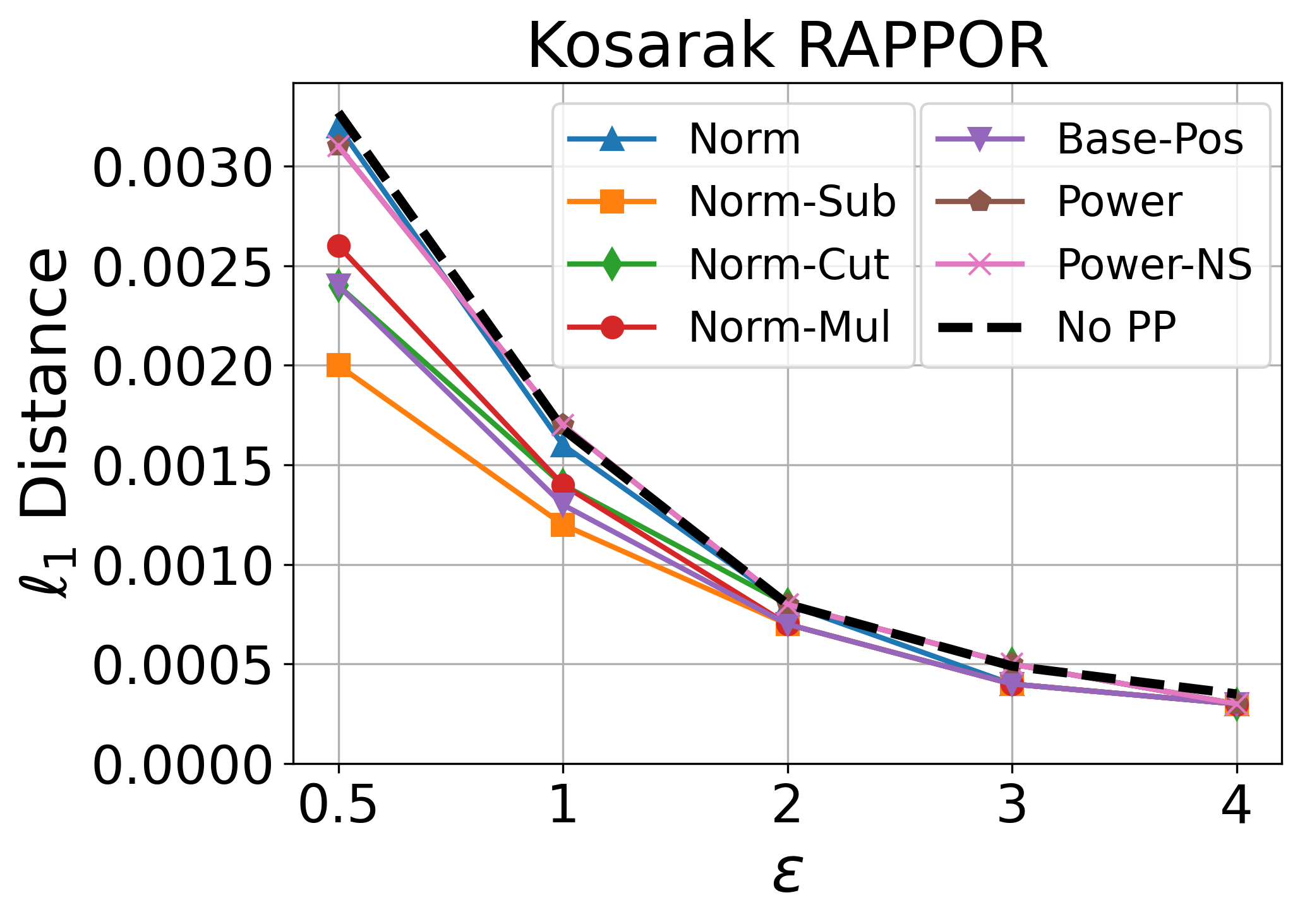}
    \end{minipage}
    \begin{minipage}[b]{0.32\textwidth}
        \centering
        \includegraphics[width=\textwidth]{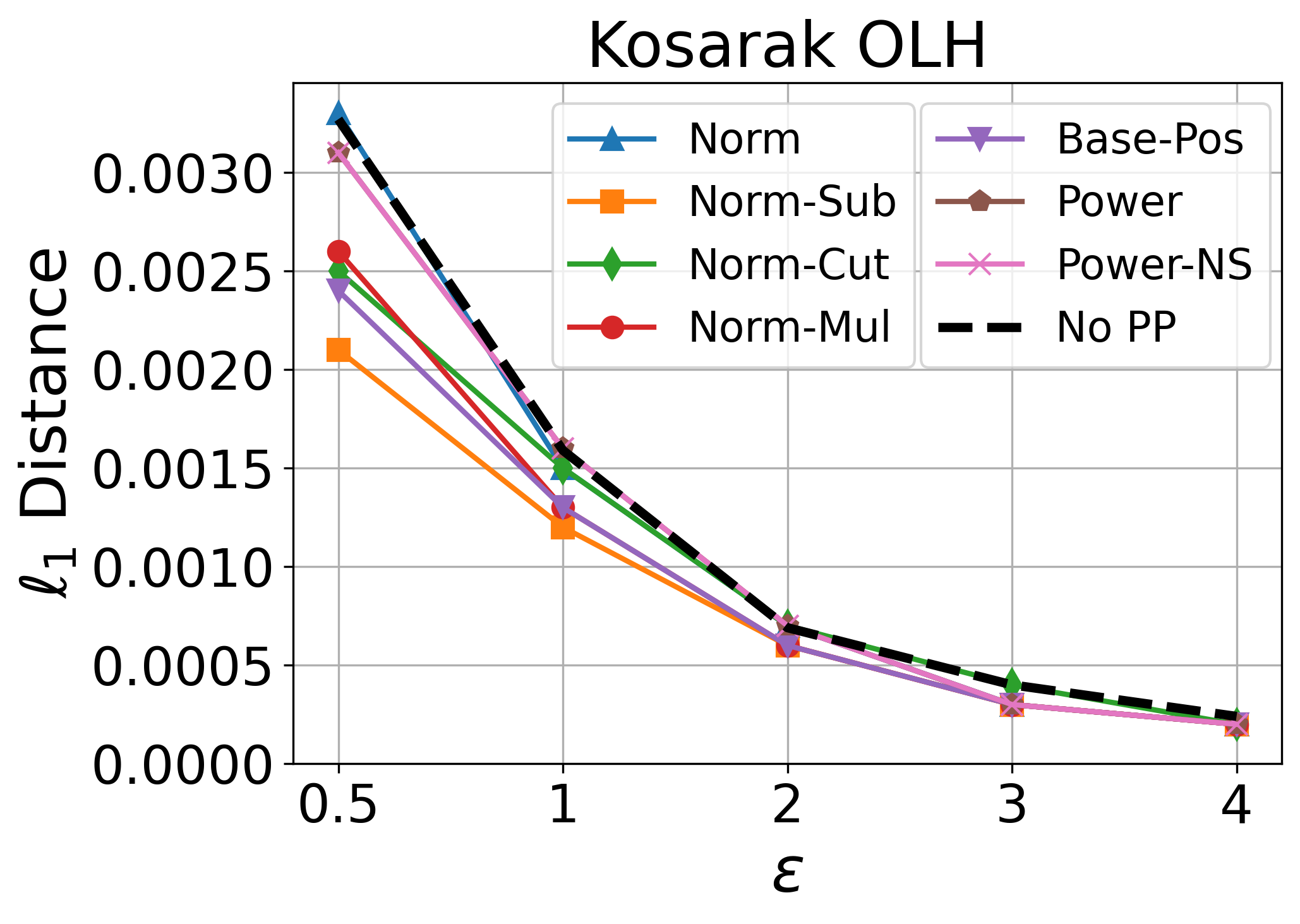}
    \end{minipage}
    \vspace{-6pt}
\caption{Average utility loss (measured via $\ell_1$ distance) of different PP methods under different datasets, protocols, and $\varepsilon$ budgets.}
\label{fig:gain}
\end{figure}

In Figure \ref{fig:gain}, we present the $\ell_1$ distance results of different PP methods under varying datasets, protocols, and $\varepsilon$ budgets. Results in this figure are the average $\ell_1$ distances over 20 runs. We observe from the plots in Figure \ref{fig:gain} that when $\varepsilon$ is smaller (such as 0.5 or 1), there exist larger differences between the $\ell_1$ distances of different PP methods. Furthermore, the benefits of using the best-performing PP method compared to not using PP (represented by the dashed line titled ``No PP'') or using a suboptimal PP method are substantial. For example, on the Adult dataset, the best-performing PP method under $\varepsilon$ = 0.5 is Norm-Mul, and its $\ell_1$ distance is almost half of No PP, Power, and Power-NS. However, as $\varepsilon$ increases, we observe that the $\ell_1$ distances of all PP methods converge, and they become similar to No PP. Considering that increased $\varepsilon$ implies reduced perturbation, these results suggest that the need for PP methods is also reduced in the case of larger $\varepsilon$ values. Since many PP methods converge as $\varepsilon$ increases, achieving the best utility also becomes less dependent on the PP method used. Combining these results, we arrive at the following take-away message. 

\begin{center}
    \setlength{\fboxrule}{0.4pt}    
    \setlength{\fboxsep}{5pt}    
    \fbox{%
        \parbox{0.9\textwidth}{%
            \centering
            PP methods can yield substantial utility benefits, especially when $\varepsilon$ is small. As $\varepsilon$ becomes larger, PP methods' utility results converge to one another, and their benefits (over not using PP) decrease.
        }%
    }
\end{center}

\begin{table}[!t]
\centering
\caption{Best-performing PP method in terms of $\ell_1$ distance under varying datasets (BMS-POS, Kosarak, Adult), LDP protocols (GRR, BLH, OLH, RAPPOR, OUE, SS), and $\varepsilon$ budgets.}
\renewcommand{\arraystretch}{1.3}
\setlength{\tabcolsep}{4pt}
\small
\begin{adjustbox}{max width=\textwidth}
\begin{tabular}{|c|c|c|c|c|c|c|c|}
\hline
\(\boldsymbol{\varepsilon}\) &  & \textbf{GRR} & \textbf{BLH} & \textbf{OLH} & \textbf{RAPPOR} & \textbf{OUE} & \textbf{SS} \\ \hline
\multirow{3}{*}{\(\varepsilon = 0.5\)} 
& BMS-POS & \cellcolor{grey-dark}Norm-Mul & \cellcolor{grey-medium}Norm-Mul & \cellcolor{grey-medium}Norm-Mul & \cellcolor{grey-medium}Norm-Mul & \cellcolor{grey-medium}Norm-Mul & \cellcolor{grey-light}Norm-Mul \\ \cline{2-8}
& Kosarak & \cellcolor{grey-dark}Norm-Sub & \cellcolor{grey-dark}Norm-Sub & \cellcolor{grey-dark}Norm-Sub & \cellcolor{grey-dark}Norm-Sub & \cellcolor{grey-dark}Norm-Sub & \cellcolor{grey-medium}Power \\ \cline{2-8}
& Adult & \cellcolor{grey-dark}Norm-Mul & \cellcolor{grey-dark}Norm-Mul & \cellcolor{grey-dark}Norm-Mul & \cellcolor{grey-dark}Norm-Mul & \cellcolor{grey-dark}Norm-Mul & \cellcolor{grey-dark}Norm-Mul \\ \hline
\multirow{3}{*}{\(\varepsilon = 1\)} 
& BMS-POS & \cellcolor{grey-dark}Norm-Mul & \cellcolor{grey-medium}Norm-Sub & \cellcolor{grey-medium}Norm-Sub & \cellcolor{grey-medium}Norm-Sub & \cellcolor{grey-medium}Norm-Sub & \cellcolor{grey-light}Norm-Sub \\ \cline{2-8}
& Kosarak & \cellcolor{grey-dark}Norm-Sub & \cellcolor{grey-dark}Norm-Sub & \cellcolor{grey-medium}Norm-Sub & \cellcolor{grey-dark}Norm-Sub & \cellcolor{grey-medium}Norm-Sub & \cellcolor{grey-medium}Norm-Sub \\ \cline{2-8}
& Adult & \cellcolor{grey-dark}Norm-Mul & \cellcolor{grey-dark}Norm-Mul & \cellcolor{grey-dark}Norm-Mul & \cellcolor{grey-medium}Norm-Mul & \cellcolor{grey-medium}Norm-Mul & \cellcolor{grey-dark}Norm-Mul \\ \hline
\multirow{3}{*}{\(\varepsilon = 2\)}
& BMS-POS & \cellcolor{grey-light}Norm-Sub & \cellcolor{grey-medium}Norm-Sub & \cellcolor{grey-light}Norm-Sub & \cellcolor{grey-medium}Norm-Sub & \cellcolor{grey-light}Norm-Mul & \cellcolor{grey-medium}Norm-Mul \\ \cline{2-8}
& Kosarak & \cellcolor{grey-dark}Norm-Sub & \cellcolor{grey-dark}Norm-Sub & \cellcolor{grey-medium}Norm-Sub & \cellcolor{grey-medium}Norm-Sub & \cellcolor{grey-medium}Norm-Sub & \cellcolor{grey-medium}Norm-Sub \\ \cline{2-8}
& Adult & \cellcolor{grey-dark}Norm-Mul & \cellcolor{grey-light}Norm-Mul & \cellcolor{grey-light}Norm-Mul & \cellcolor{grey-medium}Norm-Mul & \cellcolor{grey-light}Norm-Mul & \cellcolor{grey-light}Power \\ \hline
\multirow{3}{*}{\(\varepsilon = 3\)}
& BMS-POS & \cellcolor{grey-medium}Norm-Mul & \cellcolor{grey-light}Norm-Mul & \cellcolor{grey-light}Norm-Mul & \cellcolor{grey-light}Norm-Mul & \cellcolor{grey-light}Norm-Sub & \cellcolor{grey-medium}Norm-Mul \\ \cline{2-8}
& Kosarak & \cellcolor{grey-light}Norm-Sub & \cellcolor{grey-medium}Norm-Sub & \cellcolor{grey-light}Base-Pos & \cellcolor{grey-light}Norm-Sub & \cellcolor{grey-medium}Norm-Sub & \cellcolor{grey-light}Norm-Sub \\ \cline{2-8}
& Adult & \cellcolor{grey-medium}Power & \cellcolor{grey-light}Norm-Mul & \cellcolor{grey-medium}Norm-Sub & \cellcolor{grey-light}Norm-Mul & \cellcolor{grey-light}Norm-Sub & \cellcolor{grey-medium}Power \\ \hline
\multirow{3}{*}{\(\varepsilon = 4\)}
& BMS-POS & \cellcolor{grey-medium}Norm-Mul & \cellcolor{grey-light}Norm-Sub & \cellcolor{grey-light}Base-Pos & \cellcolor{grey-light}Norm-Sub & \cellcolor{grey-light}Norm-Mul & \cellcolor{grey-light}Norm-Mul \\ \cline{2-8}
& Kosarak & \cellcolor{grey-light}Base-Pos & \cellcolor{grey-light}Norm-Sub & \cellcolor{grey-light}Norm & \cellcolor{grey-light}Norm-Sub & \cellcolor{grey-light}Norm & \cellcolor{grey-light}Norm-Cut \\ \cline{2-8}
& Adult & \cellcolor{grey-medium}Power & \cellcolor{grey-medium}Norm-Mul & \cellcolor{grey-light}Norm-Sub & \cellcolor{grey-light}Norm-Sub & \cellcolor{grey-light}Norm-Sub & \cellcolor{grey-light}Power \\ \hline
\end{tabular}
\end{adjustbox}
\label{tab:ppm_l1}
\end{table}

In Table \ref{tab:ppm_l1}, we provide results regarding which PP method performs best under varying datasets, $\varepsilon$ budgets, and LDP protocols. The name of the best-performing PP method is written in each corresponding cell. Note that the cells are colored. We use darker colors to indicate how frequently, across 20 experiment runs, the best-performing PP method ends up being the best performer. For example, the cells corresponding to GRR and $\varepsilon$ = 0.5 are quite dark for all three datasets. This means that across 20 experiment runs, the PP methods written inside the cells (Norm-Mul and Norm-Sub) were the best performers in close to 20 of the runs. In contrast, the cells corresponding to OLH, RAPPOR, OUE, and SS with $\varepsilon$ = 4 are light-colored. This means that across 20 experiment runs, the PP methods written inside those cells were the best performers in fewer runs -- although they are still the best, other PP methods perform closely. 

We observe from Table \ref{tab:ppm_l1} that Norm-Sub and Norm-Mul usually perform better than others, especially when $\varepsilon$ = 0.5, 1, and 2. Yet, they are not always the best. Power, Base-Pos, Norm, and others can be preferred under different settings. Thus, there is no single PP method that beats others in all settings. Furthermore, we observe that when $\varepsilon$ is low (such as 0.5), the best PP method is relatively clear and consistent. Norm-Mul emerges as the best method for BMS-POS and Adult datasets, whereas Norm-Sub emerges as the best method for the Kosarak dataset. Considering the levels of darkness in the cells, we can conclude that both Norm-Mul and Norm-Sub are consistently the best performers. However, as $\varepsilon$ gradually increases, two important trends emerge. First, the dominance of a single PP method starts to diminish. A single method does not outperform all others across different LDP protocols, i.e., different LDP protocols begin to favor different PP methods. This shift indicates that, as privacy levels increase, the preference of PP methods may become more protocol-dependent. Second, with larger values of $\varepsilon$, the frequency with which the best PP method produces the lowest $\ell_1$ distance significantly decreases. It can be observed that there are many more dark-colored cells when $\varepsilon$ = 0.5 compared to much lighter-colored cells when $\varepsilon$ = 4. This indicates that with larger values of $\varepsilon$, the effectiveness of PP methods becomes more variable, making it challenging to find a consistent PP method that performs better in a given setting.

We also highlight that the results obtained in Table \ref{tab:ppm_l1} agree with the results in Figure \ref{fig:gain} in terms of which PP method performs best. Also, the convergence trend of different PP methods in Figure \ref{fig:gain} can help explain the lighter-colored cells in Table \ref{tab:ppm_l1} when $\varepsilon$ = 3 or 4. In Figure \ref{fig:gain}, we found that PP methods produce similar $\ell_1$ distances when $\varepsilon$ is large. Thus, the fact that the best-performing PP method changes frequently and the best-performing method is usually inconsistent in Table \ref{tab:ppm_l1} can be explained by this finding from Figure \ref{fig:gain}. Considering all results and discussion in this subsection, we can arrive at the following take-away message.

\begin{center}
    \setlength{\fboxrule}{0.4pt}    
    \setlength{\fboxsep}{5pt}    
    \fbox{%
        \parbox{0.9\textwidth}{%
            \centering
            Although Norm-Sub and Norm-Mul typically perform well, the best-performing PP method can be different under different protocols, $\varepsilon$ values, and datasets. The best-performing PP method is more consistent when $\varepsilon$ is small, but becomes inconsistent as $\varepsilon$ increases. 
        }%
    }
\end{center}

\subsection{Impacts of Data Characteristics}

In this section, we analyze the impacts of two data-related characteristics: data distribution and domain size. To measure the impacts of the data distribution, we utilize the Gaussian and Uniform datasets. We generate Gaussian datasets with varying standard deviations (\(\sigma = 1\), \(5\), \(10\), and \(20\)) to experiment with varying degrees of non-uniformity in $f(v)$, and a Uniform dataset which in which all $f(v)$ are equal. Smaller $\sigma$ means the data distribution is more concentrated around the mean, and therefore there is higher non-uniformity.

\begin{table}[!t]
\centering
\caption{Best-performing PP method under varying Gaussian and Uniform datasets.}
\renewcommand{\arraystretch}{1.3} 
\setlength{\tabcolsep}{6pt} 
\small 
\begin{adjustbox}{max width=\textwidth}
\begin{tabular}{|c|c|c|c|c|c|c|}
\hline
\multirow{2}{*}{} & \multirow{2}{*}{\textbf{Protocols}} & \multicolumn{5}{c|}{\textbf{Standard Deviation}} \\ \cline{3-7}
& & \(\boldsymbol{\sigma = 1}\) & \(\boldsymbol{\sigma = 5}\) & \(\boldsymbol{\sigma = 10}\) & \(\boldsymbol{\sigma = 20}\) & \textbf{Uniform} \\ \hline
\multirow{6}{*}{\textbf{\(\boldsymbol{\varepsilon} = 1\)}} 
& GRR     & \cellcolor{grey-dark}Norm-Cut & \cellcolor{grey-dark}Norm-Cut & \cellcolor{grey-dark}Norm-Mul & \cellcolor{grey-dark}Norm-Mul & \cellcolor{grey-dark}Norm-Mul \\ \cline{2-7}
& BLH     & \cellcolor{grey-dark}Norm-Cut & \cellcolor{grey-dark}Norm-Cut & \cellcolor{grey-medium}Norm-Cut & \cellcolor{grey-medium}Norm-Mul & \cellcolor{grey-medium}Norm-Mul \\ \cline{2-7}
& OLH     & \cellcolor{grey-dark}Norm-Cut & \cellcolor{grey-dark}Norm-Cut & \cellcolor{grey-light}Norm-Cut & \cellcolor{grey-medium}Norm-Mul & \cellcolor{grey-medium}Norm-Mul \\ \cline{2-7}
& RAPPOR  & \cellcolor{grey-dark}Norm-Cut & \cellcolor{grey-dark}Norm-Cut & \cellcolor{grey-light}Norm-Cut & \cellcolor{grey-medium}Norm-Mul & \cellcolor{grey-medium}Norm-Mul \\ \cline{2-7}
& OUE     & \cellcolor{grey-dark}Norm-Cut & \cellcolor{grey-medium}Norm-Cut & \cellcolor{grey-medium}Norm-Sub & \cellcolor{grey-medium}Norm-Mul & \cellcolor{grey-medium}Norm-Mul \\ \cline{2-7}
& SS      & \cellcolor{grey-dark}Norm-Cut & \cellcolor{grey-dark}Norm-Cut & \cellcolor{grey-light}Norm-Sub & \cellcolor{grey-dark}Norm-Mul & \cellcolor{grey-dark}Norm-Mul \\ \hline
\end{tabular}
\end{adjustbox}
\label{tab:gaussian}
\end{table}

The results of this experiment are shown in Table \ref{tab:gaussian}. We use the same representation and color coding as in Table \ref{tab:ppm_l1}. The $\varepsilon$ value is 1. We observe from Table \ref{tab:gaussian} that Norm-Cut is consistently the best-performing PP method for smaller $\sigma$ values such as $\sigma$ = 1 and 5. However, as $\sigma$ increases to 10 and 20, PP methods such as Norm-Mul and Norm-Sub start becoming more preferable. For $\sigma$ = 20 and the Uniform dataset, Norm-Mul becomes the best choice consistently. This progression reveals a clear trend: as non-uniformity changes, the best-performing PP method changes. Norm-Cut is the better choice when the data is non-uniform, whereas Norm-Mul becomes the better choice when the data is more uniform. This observation can be explained via the mathematical definitions of the PP methods as follows. When the data is uniform, $f(v)$ of all $v \in \mathcal{D}$ are similar and there do not exist values with substantially lower frequencies than others. In this case, the multiplicative correction applied by Norm-Mul is sufficient to ensure consistency and reduce error. On the other hand, when data is non-uniform, there exist some values $v$ such that $f(v)$ is 0 or close to 0. After LDP, some of these values end up with non-zero but low $\hat{f}(v)$. Norm-Cut applies the rule $\tilde{f}(v)  = 0$ if $\hat{f}(v) \leq \theta$ to them, i.e., these values' frequencies are post-processed to 0. Thus, it reduces error especially in datasets with high non-uniformity. 

\begin{table}[!t]
\centering
\caption{Best-performing PP method under Zipfian datasets with varying $|\mathcal{D}|$.}
\renewcommand{\arraystretch}{1.3} 
\setlength{\tabcolsep}{6pt} 
\small 
\begin{adjustbox}{max width=\textwidth}
\begin{tabular}{|c|c|c|c|c|c|c|c|c|}
\hline
\multirow{2}{*}{} & \multirow{2}{*}{\textbf{Protocols}} & \multicolumn{7}{c|}{\textbf{Domain Size $|\mathcal{D}|$}} \\ \cline{3-9}
& & \textbf{32} & \textbf{64} & \textbf{128} & \textbf{256} & \textbf{512} & \textbf{1024} & \textbf{2048} \\ \hline
\multirow{6}{*}{\(\boldsymbol{\varepsilon}\) = 1} 
& GRR     & \cellcolor{grey-light}Base-Pos & \cellcolor{grey-dark}Norm-Sub & \cellcolor{grey-dark}Norm-Sub & \cellcolor{grey-dark}Norm-Sub & \cellcolor{grey-medium}Norm-Cut & \cellcolor{grey-dark}Norm-Cut & \cellcolor{grey-dark}Norm-Cut \\ \cline{2-9}
& BLH     & \cellcolor{grey-light}Norm      & \cellcolor{grey-light}Norm-Sub & \cellcolor{grey-dark}Norm-Sub & \cellcolor{grey-dark}Norm-Sub & \cellcolor{grey-dark}Norm-Sub & \cellcolor{grey-light}Norm-Sub & \cellcolor{grey-dark}Norm-Cut \\ \cline{2-9}
& OLH     & \cellcolor{grey-medium}Norm      & \cellcolor{grey-light}Norm-Sub & \cellcolor{grey-dark}Norm-Sub & \cellcolor{grey-dark}Norm-Sub & \cellcolor{grey-dark}Norm-Sub & \cellcolor{grey-light}Norm-Cut & \cellcolor{grey-dark}Norm-Cut \\ \cline{2-9}
& RAPPOR  & \cellcolor{grey-light}Norm      & \cellcolor{grey-medium}Norm-Sub & \cellcolor{grey-dark}Norm-Sub & \cellcolor{grey-dark}Norm-Sub & \cellcolor{grey-dark}Norm-Sub & \cellcolor{grey-light}Norm-Cut & \cellcolor{grey-dark}Norm-Cut \\ \cline{2-9}
& OUE     & \cellcolor{grey-light}Norm      & \cellcolor{grey-light}Norm-Sub & \cellcolor{grey-dark}Norm-Sub & \cellcolor{grey-dark}Norm-Sub & \cellcolor{grey-dark}Norm-Sub & \cellcolor{grey-light}Norm-Cut & \cellcolor{grey-dark}Norm-Cut \\ \cline{2-9}
& SS      & \cellcolor{grey-light}Norm-Cut & \cellcolor{grey-medium}Norm-Mul & \cellcolor{grey-dark}Norm-Sub & \cellcolor{grey-medium}Power     & \cellcolor{grey-dark}Norm-Sub & \cellcolor{grey-medium}Norm-Cut & \cellcolor{grey-dark}Norm-Cut \\ \hline
\end{tabular}
\end{adjustbox}
\label{tab:zipfian}
\end{table}

In our second experiment, we analyze the impact of domain size $|\mathcal{D}|$. For this experiment, we generate Zipfian datasets with varying $|\mathcal{D}|$ = 32, 64, 128, 256, 512, 1024, and 2048. We conduct a similar analysis to Tables \ref{tab:ppm_l1} and \ref{tab:gaussian}. The results are shown in Table \ref{tab:zipfian}. For the smallest domain size $|\mathcal{D}|$ = 32, different protocols may favor different PP methods and the cells are not dark-colored, suggesting that a single PP method does not consistently outperform the others. As $|\mathcal{D}|$ increases to 64, 128, and 256, Norm-Sub becomes more dominant for most protocols. When $|\mathcal{D}|$ is further increased to 1024 and 2048, Norm-Cut emerges as the best-performing method. For the largest domain size $|\mathcal{D}|$ = 2048, Norm-Cut becomes the best-performing protocol consistently. The superior performance of Norm-Sub and Norm-Cut for larger $|\mathcal{D}|$ agrees with results from Table \ref{tab:gaussian}, since both Norm-Sub and Norm-Cut post-process negative frequencies and low frequencies to 0. When $|\mathcal{D}|$ is large, as $|\mathcal{P}|$ is constant, there is a higher likelihood of obtaining low-frequency values $v \in \mathcal{D}$, whose frequencies may remain low or become negative after LDP perturbation. Norm-Sub and Norm-Cut help in reducing the errors in these cases. Overall, the results also demonstrate that the best PP method changes when $|\mathcal{D}|$ is changed, e.g., multiple PP methods perform well for small domain sizes, Norm-Sub emerges as the best approach for medium-sized domains (128 to 512), and Norm-Cut is the best for large domains ($>$ 1024). Considering the results and discussion in this subsection, we arrive at the following take-away message.

\begin{center}
    \setlength{\fboxrule}{0.4pt}    
    \setlength{\fboxsep}{5pt}    
    \fbox{%
        \parbox{0.9\textwidth}{%
            \centering
            The best-performing PP method is dependent on data characteristics, e.g., data distribution (non-uniformity) and domain size.
        }%
    }
\end{center}

\subsection{Impacts of the Utility Metrics} 

In the experiments reported so far, $\ell_1$ distance was used as the utility metric. In this section, we examine how the choice of utility metric impacts the performance of PP methods. We use all four of the utility metrics described in Section \ref{sec:UtilityMetrics}: $\ell_1$ distance, $\ell_2$ distance, KL-divergence, and EMD. We report experiments with varying protocols, but fixed $\varepsilon$ = 1 and the Kosarak dataset. The results are given in Table \ref{tab:utility}. The same format and color coding as the previous tables are used. 

\begin{table}[!t]
\centering
\caption{Best-performing PP method under varying utility metrics. Kosarak dataset and $\varepsilon$ = 1 are used.}
\renewcommand{\arraystretch}{1.3} 
\setlength{\tabcolsep}{6pt} 
\small 
\resizebox{\textwidth}{!}{
\begin{tabular}{|c|c|c|c|c|c|}
\hline
\multirow{2}{*}{} & \multirow{2}{*}{\textbf{Protocols}} & \multicolumn{4}{c|}{\textbf{Utility Metric}} \\ \cline{3-6}
& & \textbf{$\ell_1$ distance} & \textbf{$\ell_2$ distance} & \textbf{KL-divergence} & \textbf{EMD} \\ \hline
\multirow{6}{*}{\(\boldsymbol{\varepsilon}\) = 1} 
& GRR     & \cellcolor{grey-dark}Norm-Sub  & \cellcolor{grey-dark}Norm-Sub & \cellcolor{grey-dark}Norm-Sub & \cellcolor{grey-medium}Norm  \\ \cline{2-6}
& BLH     & \cellcolor{grey-dark}Norm-Sub  & \cellcolor{grey-dark}Norm-Sub  & \cellcolor{grey-dark}Norm-Sub & \cellcolor{grey-light}Norm  \\ \cline{2-6}
& OLH     & \cellcolor{grey-dark}Norm-Sub   & \cellcolor{grey-dark}Norm-Sub   & \cellcolor{grey-light}Norm-Sub  & \cellcolor{grey-light}Norm-Sub \\ \cline{2-6}
& RAPPOR  & \cellcolor{grey-dark}Norm-Sub  & \cellcolor{grey-dark}Norm-Sub  & \cellcolor{grey-medium}Norm-Sub  & \cellcolor{grey-light}Norm-Sub \\ \cline{2-6}
& OUE     & \cellcolor{grey-medium}Norm-Sub  & \cellcolor{grey-dark}Norm-Sub   & \cellcolor{grey-light}Norm-Sub  & \cellcolor{grey-light}Norm-Sub \\ \cline{2-6}
& SS      & \cellcolor{grey-dark}Norm-Sub  & \cellcolor{grey-dark}Norm-Sub  & \cellcolor{grey-dark}Power  & \cellcolor{grey-light}Power-NS  \\ \hline
\end{tabular}
}
\label{tab:utility}
\end{table}

For most protocols, $\ell_1$ distance and $\ell_2$ distance metrics consistently favor Norm-Sub. This observation underscores that the $\ell_1$ and $\ell_2$ distance metrics, which are similar in nature, agree in terms of the favorable PP method. In contrast, KL-divergence exhibits slightly more variability compared to $\ell_1$ and $\ell_2$. While Norm-Sub still performs well for most protocols, there are notable deviations. For example, for SS, Power is consistently the best-performing PP method. The reduced darkness of the cells corresponding to KL-divergence further highlights the varying nature of KL-divergence in comparison to $\ell_1$ and $\ell_2$ distance. On the other hand, EMD diverges even more significantly from $\ell_1$ and $\ell_2$, favoring a more distributed selection of methods. Norm-Sub still remains a leading contender but with substantially reduced consistencies, as seen in protocols like OLH, RAPPOR, and OUE. In the remaining protocols (GRR, BLH, SS), alternative PP methods such as Norm and Power-NS perform the best.  

Overall, the results indicate that the best-performing PP method is indeed influenced by the choice of utility metric. Metrics that are inherently similar to one another tend to prefer the same PP methods, whereas different metrics (such as EMD) tend to prefer different PP methods.

\begin{center}
    \setlength{\fboxrule}{0.4pt}    
    \setlength{\fboxsep}{5pt}    
    \fbox{%
        \parbox{0.9\textwidth}{%
            \centering
            Norm-Sub typically performs well across multiple utility metrics, but the optimal PP method can change according to the choice of metric.
        }%
    }
\end{center}

\section{LDP$^3$: An Open-Source Benchmark Platform} \label{sec:LDPCube}

The results of our experimental analysis show that the optimal choice of PP method depends on multiple factors, including the LDP protocol, $\varepsilon$ budget, data characteristics (e.g., statistical distribution, domain size), and utility metric. When these factors are different, the best-performing PP method is also different. This motivated the following question: Can we design a practical system that helps a researcher or practitioner choose the best PP method for their setting, i.e., for their choice of data, protocol, budget, and utility metric? 

We designed and developed LDP$^3$ to address this need (pronounced LDP-Cube, stands for: \textbf{L}ocal \textbf{D}ifferential \textbf{P}rivacy with \textbf{P}ost \textbf{P}rocessing). LDP$^3$ is an open-source benchmark platform available on Github\footnote{\url{https://github.com/alrzakh/LDPcube}}. It contains implementations of 6 LDP protocols, 7 PP methods, 4 utility metrics, and 6 datasets in a multi-threaded setup. It aims to offer a comprehensive and extensible platform for researchers and practitioners to evaluate combinations of different LDP protocols and PP methods using their desired datasets, privacy budgets, and utility metrics. The results obtained from LDP$^3$ can assist researchers and practitioners in choosing the best-performing PP method in their custom settings. 

\begin{figure}[!t]
    \centering
    \includegraphics[width=.7\textwidth]{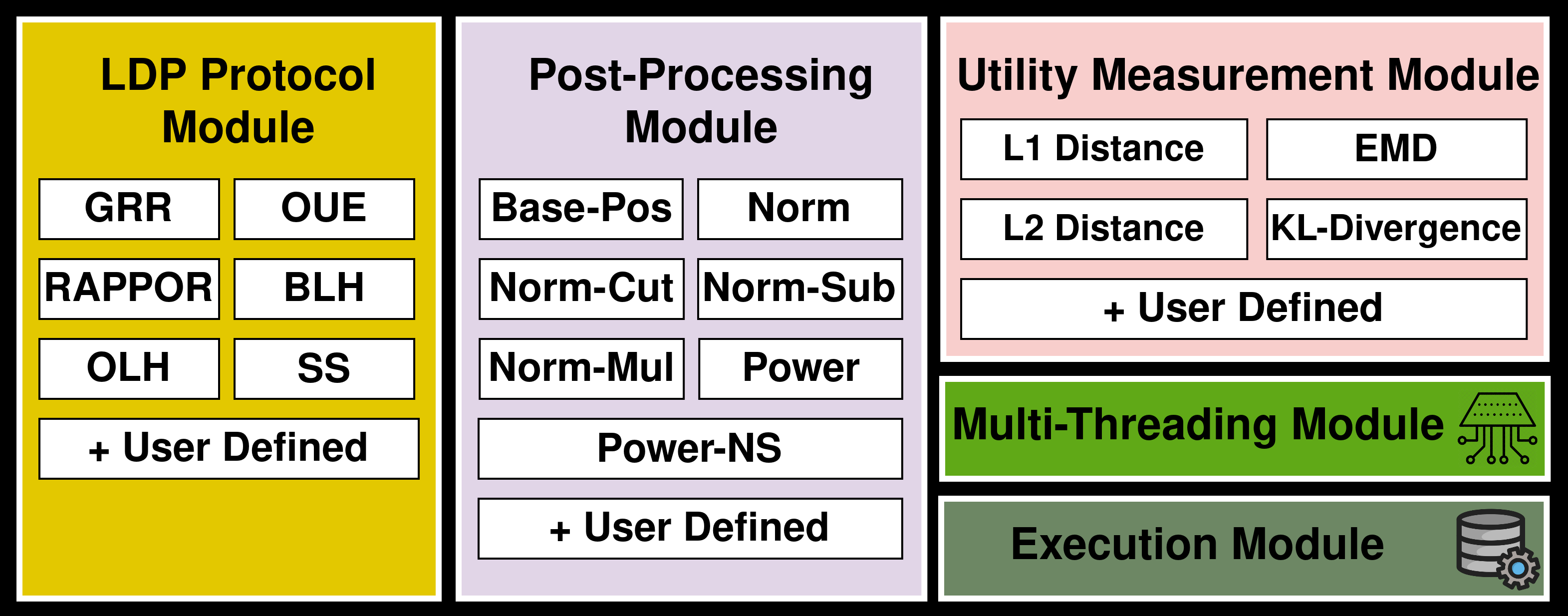}
    \vspace{-4pt}
\caption{LDP$^3$ system architecture.}
\label{fig:ldpcube}
\vspace{-2pt}
\end{figure}

The architectural design of LDP$^3$ is shown in Figure \ref{fig:ldpcube}. As it can be observed from Figure \ref{fig:ldpcube}, LDP$^3$ consists of the following modules:
\begin{itemize}
    \item The LDP Protocol Module contains implementations of 6 state-of-the-art LDP protocols: GRR, BLH, OLH, RAPPOR, OUE, and SS. It is extensible to accommodate new protocols, e.g., protocols can be added by implementing the functionalities for user-side perturbation and server-side estimation.
    \item The Post-Processing Module contains implementations of 7 PP methods explained in Section \ref{sec:PPmethods}. All methods take as input the estimated frequencies $\hat{f}(v)$ and produce post-processed $\tilde{f}(v)$. New PP methods can also be added in the future.
    \item The Utility Measurement Module contains implementations of utility metrics to measure the differences between $f(v)$ and $\tilde{f}(v)$. Currently, the 4 utility metrics explained in Section \ref{sec:UtilityMetrics} are implemented in this module.
    \item The Multi-Threading Module leverages Python's concurrency features to parallelize the simulation of data perturbation and aggregation processes. It is useful when LDP$^3$ is run on multi-threaded or multi-core CPUs. In essence, it enables $\mathcal{P}$ to be divided into the number of available threads, simulates data collection for each subset of $\mathcal{P}$ in a different thread in parallel, and eventually combines the results from the different threads.
    \item Finally, the Execution Module facilitates the handling of datasets (e.g., reading or generating datasets from Section \ref{sec:Datasets}) and provides a command-line interface to run experiments.
\end{itemize}

\vspace{-4pt}
\section{Conclusion} \label{sec:Conclusion}
\vspace{-2pt}

In this paper, we performed an experimental analysis of PP methods and showed that PP methods can significantly improve utility in LDP, especially under strict privacy regimes (small $\varepsilon$). However, we also showed that this benefit is neither universal nor static: the optimal PP method depends on a combination of factors, including the LDP protocol, privacy budget, data characteristics (e.g., statistical distribution, domain size), and the specific utility metric. We then designed and developed an open-source benchmark platform for LDP protocols and PP methods called LDP$^3$, which integrates all components used in our experimental analysis. By making LDP$^3$ modular, extensible, and multi-threaded, we aimed to provide a platform that is helpful for both LDP researchers and practitioners.

There are several avenues for future work. First, we plan to integrate additional LDP protocols and PP methods into LDP$^3$ to further increase its breadth. Second, we plan to add a new module to evaluate protocols' adversarial aspects, e.g., privacy attacks and poisoning attacks. Third, we plan to design a graphical user interface (GUI) for LDP$^3$ and conduct human studies to test how convenient it is for non-experts to use LDP$^3$ for protocol and PP method selection.

\vspace{-2pt}

\subsubsection{\ackname} This study was supported by the Scientific and Technological Research Council of Türkiye (TUBITAK) under Grant Number 123E179. The authors thank TUBITAK for their support.

\bibliographystyle{splncs04}
\bibliography{references}

\end{document}